\documentclass[aps,pra,twocolumn,showpacs,amsmath,amssymb,floatfix,footinbib,superscriptaddress]{revtex4-1}
\usepackage{graphicx}
\usepackage{amsmath}
\usepackage{latexsym}
\usepackage{amsmath}
\usepackage{amssymb}
\usepackage{float}
\usepackage{epstopdf}
\usepackage{bm}
\usepackage{amsfonts}
\usepackage[ansinew]{inputenc}
\usepackage[usenames,dvipsnames]{pstricks}
\usepackage{subfigure}
\usepackage{epsfig}
\usepackage{longtable}
\usepackage{url}
\usepackage{hyperref}
\usepackage{color}
\definecolor{darkblue}{rgb}{0.0,0.0,0.3}
\hypersetup{colorlinks,breaklinks,linkcolor=blue,urlcolor=blue,anchorcolor=blue,citecolor=blue}

\begin{document}
\baselineskip=14pt
\title{Study of optical nonlinearity of a highly dispersive medium using optical heterodyne detection technique}

\author{Arup Bhowmick}
\email{arup.b@niser.ac.in}
\affiliation{School of Physical Sciences, National Institute of Science Education and Research, Bhubaneswar, 752050, India}
\author{Sushree S. Sahoo}
\affiliation{School of Physical Sciences, National Institute of Science Education and Research, Bhubaneswar, 752050, India}
\author{Ashok K Mohapatra}%
\email{a.mohapatra@niser.ac.in}
\affiliation{School of Physical Sciences, National Institute of Science Education and Research, Bhubaneswar, 752050, India}

\date{\today}

\begin{abstract}
We discuss the optical heterodyne detection technique to study the absorption and dispersion of a probe beam propagating through a medium with a narrow resonance. The technique has been demonstrated for Rydberg Electro-magnetically induced transparency (EIT) in rubidium thermal vapor and the optical non-linearity of a probe beam with variable intensity has been studied. A quantitative comparison of the experimental result with a suitable theoretical model is presented. The limitations and the working regime of the technique are discussed.

\end{abstract}

\keywords{Atom-light interaction, Rydberg EIT, Optical non-linearity}
\maketitle
\section{Introduction}
Self phase modulation (SPM) and cross phase modulation (XPM) are at the heart of strong photon-photon interactions inside a medium which
plays an important role in building quantum gates~\cite{milb89,chua95}, quantum entanglement~\cite{luki00} and non-demolition measurement~\cite{imot85} of single photons. Strong XPM of photons based on EIT has been theoretically proposed~\cite{schm96} and demonstrated in 
thermal vapors~\cite{wang01,chan04,li08} as well as in cold atoms~\cite{kang03,lo10,chen13}. Enhanced SPM of photons mediated by Rydberg blockade interaction in atomic vapor has been proposed~\cite{frie05,sevi11,gors11}. Rydberg blockade induced photon-photon interaction has been experimentally demonstrated for weak classical light~\cite{prit10,pari12} and single photons in cold 
atoms~\cite{peyr12,firs13}.

EIT based XPM has been measured using various interferometric techniques~\cite{wang01,chan04,li08,kang03,lo10}. Optical heterodyne is one of such techniques which has been extensively used for the measurement of absorption and dispersion of coherent 2-photon transition in an atomic ensemble~\cite{mull96}, Zeeman coherence induced anomalous dispersion~\cite{akul99}, and enhanced Kerr non-linearity in 2-level atoms~\cite{akul04}. The technique has also been used to measure the XPM of a probe and a control beam in an N-system using cold atoms~\cite{kang03,lo10,han08}. The basic principle of the technique is based on using two probe beams propagating through the dispersive medium with a frequency offset larger than the resonance line width. Both the beams can't be on resonance while scanning their frequencies and hence, they undergo different phase shifts. This differential phase shift appears in their beat signal which can be measured by comparing with the phase of a reference beat signal of the same two beams and gives information about the dispersion. If the probe beams are sufficiently weak, then the measured optical non-linearity using this technique can be compared with the standard models involving a single probe beam. However, if the intensity of one of the probe beams is increased, then the strong probe beam dresses the atoms interacting with the weak probe beam which leads to the erroneous measurement of the non-linearity. If both the probe beams are strong then the issue is even more serious.

In this article, we have demonstrated this technique to measure the SPM of a probe beam propagating through a Rydberg EIT medium in
rubidium thermal vapor. We show that the observed probe transmission and dispersion can't be explained with the standard EIT theory
for the probe beam with large intensity. We present a model of EIT consisting of a strong coupling beam and two probe beams with a frequency offset to explain the experimental data. The paper is organized as follows. In the next section, we discuss the theoretical model. The experimental method of heterodyne detection technique is presented in section III followed by the measurement of optical non-linearity in section IV.

\section{Theoretical Model}

\begin{figure}[t]
\epsfig{file=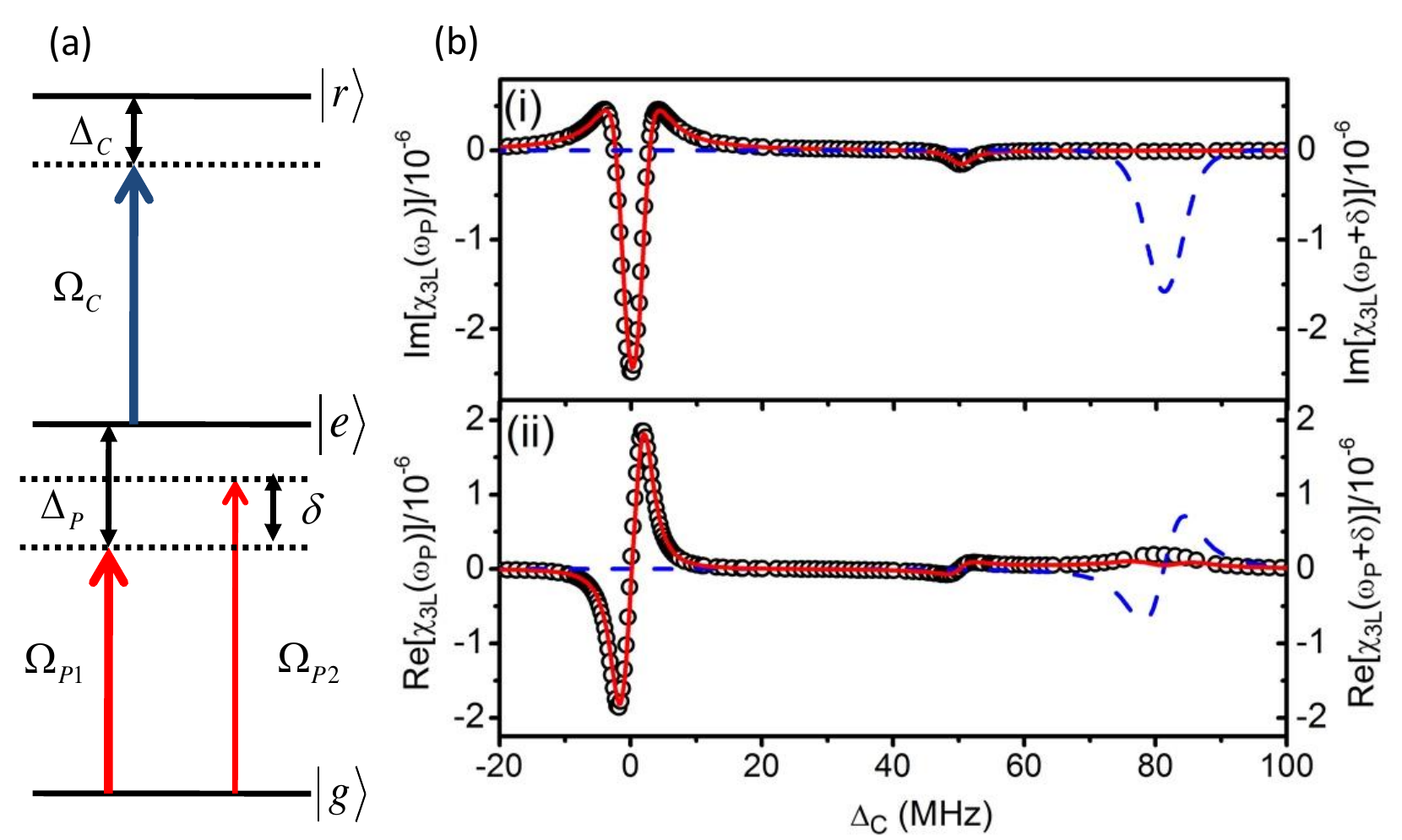,clip=,angle=0,width=9.0cm}
\caption[]{(a) Energy level diagram of EIT in ladder configuration. Two probe fields couple the transition $|g\rangle$ $\longrightarrow$ 
$|e\rangle$. The coupling laser couples the transition $|e\rangle$ $\longrightarrow$ $|r\rangle$. The probe (coupling) detuning is 
$\Delta_p\left(\Delta_c\right)$ and the frequency offset between the probe beams is $\delta$. (b) Imaginary (i) and real (ii) part of 
$\chi_{3L}(\omega_p)$ and $\chi_{3L}(\omega_p+\delta)$ as a function of coupling laser detuning. The parameters used in the model are, 
$\Delta_p=-50$ MHz, $\delta=50$ MHz, $\Omega_{p1}=5$ MHz, $\Omega_{p2}=0.5$ MHz, $\Omega_c=2.5$ MHz, $k_c=\frac{1}{480}$ nm$^{-1}$ and $k_p=\frac{1}{780}$ nm$^{-1}$. Doppler averaging was done using temperature of the vapor to be $T=300$ K. The blue dotted and red solid lines are susceptibilities of the strong and the weak probe beam respectively. The open circles show susceptibility of the weak probe beam calculated using the approximation discussed in the text.}
\label{fig1}
\end{figure}

In order to explain the transmission and dispersion of a probe beam propagating through Rydberg EIT medium, we consider a model of three-level atomic system interacting with two probe laser fields and a coupling laser field in ladder configuration as shown in fig~\ref{fig1}(a). The coupling laser field with frequency $\omega_c$ counter-propagates the co-propagating probe beams with frequencies $\omega_p$ and $\omega_p+\delta$ through the vapor cell. The probe field with frequency $\omega_p+\delta$ is considered as a weak field. In a suitable rotating frame and with rotating wave approximation (RWA), the total Hamiltonian of the system can be written as,
\begin{displaymath}
\mathbf{H} =\frac{\hbar}{2}
 \left( \begin{array}{ccc}
0& \Omega^{*}_{p1} +\Omega^{*}_{p2} e^{i\delta t} & 0 \\
\Omega_{p1}+\Omega_{p2} e^{-i\delta t} & 2(\Delta_{p}-k_{p}v) & \Omega^{*}_{c} \\
0 & \Omega_{c} & 2(\Delta_{2}-\Delta k v) \\
\end{array} \right).
\end{displaymath}

where $\Omega_{p1}$, $\Omega_{p2}$ and $\Omega_{c}$ are the Rabi frequencies of the strong probe, weak probe and the coupling beams, respectively. $k_{c}$ and $k_{p}$ are the wave vectors of the coupling and probe lasers with $\Delta k=k_p-k_c$. The two photon detuning is given by, $\Delta_{2}=\Delta_{p} + \Delta_{c}$. $v$ is the velocity of the atoms in the vapor. The density matrix equation is given by, 
$\dot{\rho}=\frac{-i}{\hbar}[H,\rho]+\mathcal{L}_{D}(\rho)$. $\mathcal{L}_{D}(\rho)$ is the Lindblad operator which takes care of the decoherences 
in the system. The population decay rate of the channels, $|r\rangle \rightarrow |e\rangle$ is denoted by $\Gamma_{re}$ and $|e\rangle \rightarrow |g\rangle$ is denoted by $\Gamma_{eg}$. Due to the finite transit time of the thermal atoms through the cross section of the beams, we include the population decay rate of the Rydberg state to the ground state as $\Gamma_{rg}$. In our model, the decay time scales used 
are $\Gamma_{re}=10$ kHz, $\Gamma_{eg}=6$ MHz and $\Gamma_{rg}=200$ kHz.

The steady state density matrix equations are solved perturbatively. A similar approach is used to calculate the 4-wave mixing in 2-level atoms as discussed in reference~\cite{boyd08}. The density matrix of the system can be expanded as, $\rho = \rho^{(0)}+\rho^{(1)} 
e^{-i\delta t}+\rho^{(-1)} e^{i\delta t}$ and substituted in the density matrix equations. Equating the coefficients of 
$e^{-i\delta t}$ with $\delta=0$ gives the zeroth order equations.
\begin{eqnarray}
&&\frac{\Omega_{p1}}{2}\rho^{(0)}_{eg} - \frac{\Omega^{*}_{p1}}{2}\rho^{(0)}_{ge}-i\Gamma_{eg}\rho^{(0)}_{ee}
-i\Gamma_{rg}\rho^{(0)}_{rr}=0 \\
&&\frac{\Omega^{*}_{c}}{2}\rho^{(0)}_{er}-\frac{\Omega_{c}}{2}\rho^{(0)}_{re}+i\Gamma_{2}\rho^{(0)}_{rr}=0 \\
&&\left(\Delta_{p}-k_pv-i\frac{\Gamma_{eg}}{2}\right)\rho^{(0)}_{ge}-\frac{\Omega_{p1}}{2}(2\rho^{(0)}_{ee}\rho^{(0)}_{rr}-1) \nonumber\\
&&+\>\frac{\Omega^{*}_{c}}{2}\rho^{(0)}_{gr}=0 \\
&&\left(\Delta_2-\Delta kv-i\frac{\Gamma_2}{2}\right)\rho^{(0)}_{gr}-\frac{\Omega_{p1}}{2}\rho^{(0)}_{er}\nonumber\\
&&+\>\frac{\Omega_{c}}{2}\rho^{(0)}_{ge}=0 \\
&&\left(\Delta_c+k_cv-i\frac{\Gamma_3}{2}\right)\rho^{(0)}_{er}-\frac{\Omega_{c}}{2}(\rho^{(0)}_{rr}+\rho^{(0)}_{ee}) \nonumber\\
&&+\>\frac{\Omega^{*}_{p1}}{2}\rho^{(0)}_{gr}=0 \\
\nonumber
\end{eqnarray}
where $\Gamma_{2}=\Gamma_{re}+\Gamma_{rg}$ and $\Gamma_{3}=\Gamma_{eg}+\Gamma_{re}+\Gamma_{rg}$.
The zeroth order equations are same as the equations of EIT for the probe beam with Rabi frequency $\Omega_{p1}$ and can be solved exactly. 
Equating the coefficients of $e^{-i\delta t}$ gives the first order equations which can be solved if the 2nd order terms are neglected. 
Hence, the model is valid if one of the probe beams is weak. The first order equations in the steady state are given by,
\begin{widetext}
\begin{eqnarray}
&&\left(\Delta_p+\delta-k_pv+i\frac{\Gamma_{eg}}{2}\right)\rho^{(1)}_{eg}+\frac{\Omega_{c}}{2}\rho^{(1)}_{rg}-\frac{\Omega_{p1}}{2}\left(2\rho^{(1)}_{ee}+\rho^{(1)}_{rr}\right)-\frac{\Omega_{p2}}{2}\left(2\rho^{(0)}_{ee}+\rho^{(0)}_{rr}-1\right)=0 \\
&&\left(\Delta_2+\delta-\Delta kv+i\frac{\Gamma_{2}}{2}\right)\rho^{(1)}_{rg}+\frac{\Omega_{c}}{2}\rho^{(1)}_{eg}-\frac{\Omega_{p1}}{2}\rho^{(1)}_{re}
-\frac{\Omega_{p2}}{2}\rho^{(0)}_{re}=0 \\
&&\left(\Delta_c+\delta+k_cv+i\frac{\Gamma_3}{2}\right)\rho^{(1)}_{re}-\frac{\Omega_{p1}}{2}\rho^{(1)}_{rg}-\frac{\Omega_{c}}{2}(\rho^{(1)}_{rr}-\rho^{(1)}_{ee})=0 \\
&&\left(\Delta_p-\delta-k_pv-i\frac{\Gamma_{eg}}{2}\right)\rho^{(1)}_{ge}+\frac{\Omega_{c}}{2}\rho^{(1)}_{gr}-\frac{\Omega_{p1}}{2}(2\rho^{(1)}_{ee}+\rho^{(1)}_{rr})=0 \\
&&\left(\Delta_{2}-\delta-\Delta kv-i\frac{\Gamma_2}{2}\right)\rho^{(1)}_{gr}+\frac{\Omega_{c}}{2}\rho^{(1)}_{ge}-\frac{\Omega_{p1}}{2}\rho^{(1)}_{er}=0\\
&&\left(\Delta_{c}-\delta+k_cv-i\frac{\Gamma_3}{2}\right)\rho^{(1)}_{er}+\frac{\Omega_{c}}{2}(\rho^{(1)}_{ee}-\rho^{(1)}_{rr})-\frac{\Omega_{p2}}{2}\rho^{(0)}_{gr}-\frac{\Omega_{p1}}{2}\rho^{(1)}_{gr}=0 \\
&&\frac{\Omega_{c}}{2}\left(\rho^{(1)}_{re}-\rho^{(1)}_{er}\right)-\delta\rho^{(1)}_{rr}-i\Gamma_{2}\rho^{(1)}_{rr}=0 \\
&&\frac{\Omega_{p1}}{2}\left(\rho^{(1)}_{eg}-\rho^{(1)}_{ge}\right)+\frac{\Omega_{c}}{2}\left(\rho^{(1)}_{er}-\rho^{(1)}_{re}\right)+\frac{\Omega_{p2}}{2}\rho^{(0)}_{ge}-\delta\rho^{(1)}_{ee}-i\Gamma_{eg}\rho^{(1)}_{ee}+i\Gamma_{re}\rho^{(1)}_{rr}=0 \\
\nonumber
\end{eqnarray}
\end{widetext}
Using the fact that $\rho^{*}_{ij}=\rho_{ji}$, it can be shown that $\rho^{(0)*}_{ij}=\rho^{(0)}_{ji}$ and 
$\rho^{(1)*}_{ij}=\rho^{(-1)}_{ji}$. Assuming that the system is closed and using $\rho_{gg}+\rho_{ee}+\rho_{rr}=1$, 
we get 8 independent first order equations.

\begin{figure}[t]
\epsfig{file=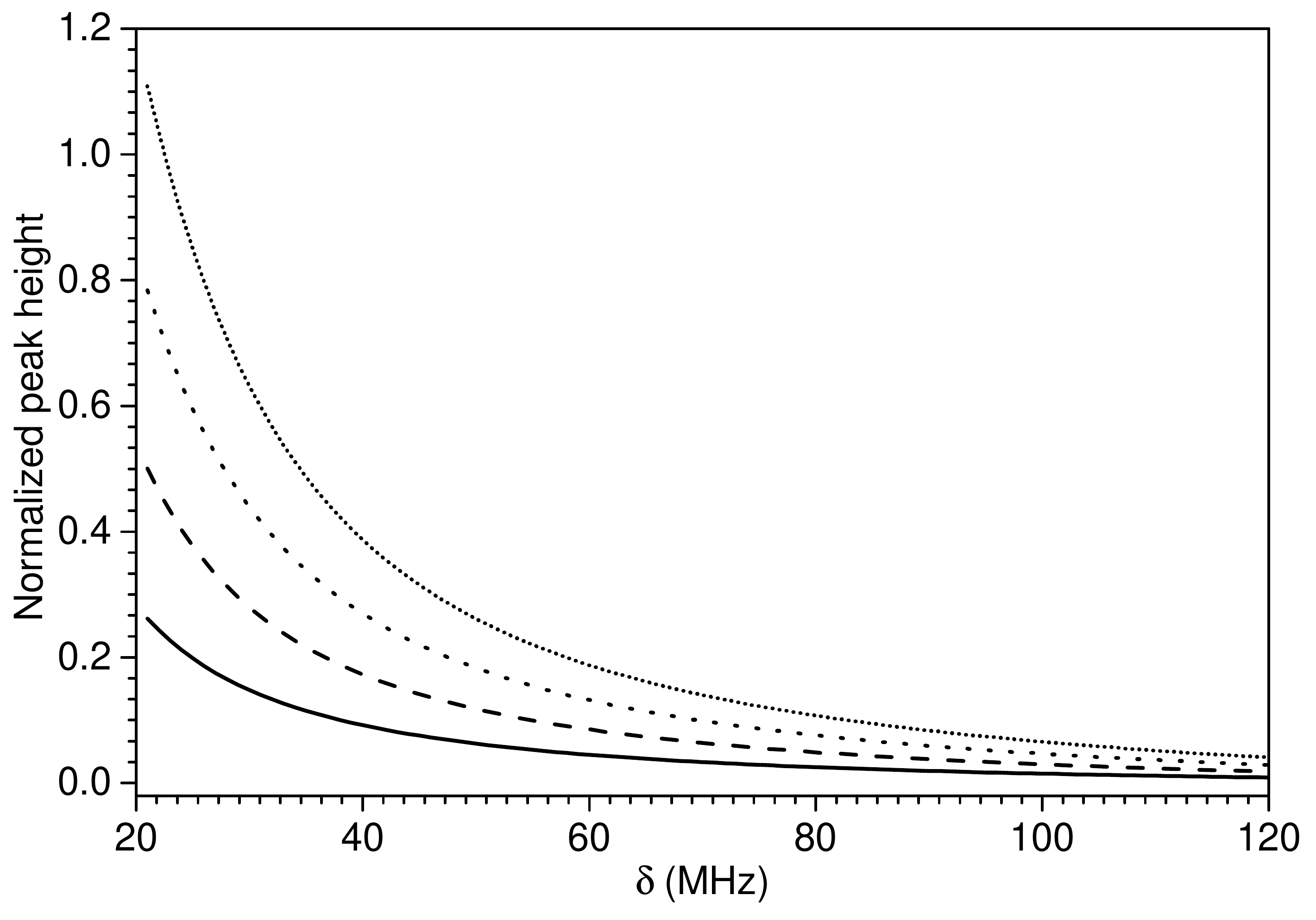,clip=,angle=0,width=9.0cm}
\caption[]{Variation of the EIT transmission peak height due to dressed atoms as a function of beat frequency. The curves are generated using the model for the coupling Rabi frequency $2.5$ MHz and for probe Rabi frequencies $4$ MHz (solid line), $6$ MHz (dashed line), 
$8$ MHz (big dotted line), and $10$ MHz (small dotted line).}
\label{fig2}
\end{figure}

Zeroth order equations are solved numerically in steady state for the zeroth order matrix elements $\rho^{(0)}_{i,j}$ $\forall$ $i$, $j$ and are substituted in the first order equations. The first order equations are then solved numerically in steady state to determine $\rho^{(1)}_{eg}$. The susceptibility of the strong probe averaged over the thermal motion of the atoms can be calculated as $\chi(\omega_p)=\frac{2N\left|\mu_{eg}\right|^2}{\epsilon_0\hbar\Omega_{p1}}\frac{1}{\sqrt{2\pi}v_p}\int^{\infty}_{-\infty}\rho^{(0)}_{eg}e^{-v^2/2v^2_p}dv$ where $v_p$ is the most probable speed of the atoms, $N$ is the density and $\mu_{eg}$ is the dipole moment of the transition $|g\rangle\longrightarrow|e\rangle$. Similarly, the susceptibility of the weak probe can be determined using $\chi(\omega_p+\delta)=\frac{2N\left|\mu_{eg}\right|^2}{\epsilon_0\hbar\Omega_{p2}}\frac{1}{\sqrt{2\pi}v_p}\int^{\infty}_{-\infty}\rho^{(1)}_{eg}e^{-v^2/2v^2_p}dv$. Heterodyne detection technique is sensitive only to the 2-photon transition and hence the susceptibility of the probe in the absence of the coupling beam can't be detected. To compare with the experiment, we define the susceptibility only due to 2-photon transition as $\chi_{3L}(\omega_p)=\chi(\omega_p)-\chi_{2L}(\omega_p)$ and $\chi_{3L}(\omega_p+\delta)=\chi(\omega_p+\delta)-\chi_{2L}(\omega_p+\delta)$, where $\chi_{2L}$ is the susceptibility of the probes in the absence of the coupling beam. $\chi_{3L}$ calculated from the model is depicted in figure (\ref{fig1}b). As shown in the figure, two EIT peaks associated with both the probe beams are observed. However, the frequency difference between the EIT peaks doesn't match with the offset frequency, but is scaled as $\frac{k_c}{k_p}\delta$. The scaling can easily be understood by looking at the EIT equations. EIT resonance peak for the strong probe is observed if $\Delta_2-\Delta kv=0$ and $\Delta_p-k_pv=0$. So EIT resonance of the strong probe appears at $\Delta_c=-\frac{k_c}{k_p}\Delta_p$. Similarly, EIT peak for the weak probe is observed if $\Delta_2+\delta-\Delta kv=0$ and $\Delta_p+\delta-k_pv=0$. Hence, EIT peak of the weak probe appears at $\Delta_{c1}=-\frac{k_c}{k_p}(\Delta_{p}+\delta)$. The spectral difference between the EIT peaks is $\Delta_{c1}-\Delta_c=\frac{k_c}{k_p}\delta$. Similar scaling of Rydberg EIT peaks associated with the hyperfine transitions in rubidium thermal vapor has been reported in reference~\cite{moha07}.   

As shown in figure (\ref{fig1}b), an unexpected small peak is observed for the weak probe susceptibility when coupling laser is detuned by 
$50$ MHz from the weak probe EIT peak. In order to get an insight of the origin of this peak, we use the following approximations to 
simplify the first order equations. Since the probe beam is weak, it cannot raise the population in the excited states. Hence, 
$\rho^{(1)}_{ee}\approx\rho^{(1)}_{rr}\approx 0$. Using this approximations, the first order equations are reduced to
\begin{eqnarray}
&&\left(\Delta_p+\delta-k_pv+i\frac{\Gamma_{eg}}{2}\right)\rho^{(1)}_{eg}+\frac{\Omega_{c}}{2}\rho^{(1)}_{rg}\nonumber\\
&&+\>\frac{\Omega_{p2}}{2}\left(2\rho_{ee}^{(0)}+\rho_{rr}^{(0)}-1\right)=0 \\
&&\left(\Delta_2+\delta-\Delta kv+i\frac{\Gamma_{2}}{2}\right)\rho^{(1)}_{rg}+\frac{\Omega_{c}}{2}\rho^{(1)}_{eg} \nonumber \\
&&-\frac{\Omega_{p1}}{2}\rho^{(1)}_{re}-\frac{\Omega_{p2}}{2}\rho^{(0)}_{re}=0 \\
&&\left(\Delta_c+\delta+k_cv+i\frac{\Gamma_3}{2}\right)\rho^{(1)}_{re}-\frac{\Omega_{p1}}{2}\rho^{(1)}_{rg}=0 
\end{eqnarray}
In the absence of the strong probe beam, $\Omega_{p1}=0$ and all the zeroth order matrix elements are equal to zero 
and equation (16) leads to $\rho^{(1)}_{re}=0$. Under this condition, it can be shown that equations (14) and (15) exactly 
give the EIT equations in weak probe limit. In the presence of the strong probe beam with frequency offset $\delta=50$ MHz, 
the extra zeroth order terms in the equations leads to the appearance of the small peak. To understand it further, let the 
weak probe interacts with the zero velocity class of atoms. So, the main EIT peak of the weak probe appears at $\Delta_c=0$. 
The presence of the strong probe dresses the same zero velocity class of atoms which are excited to the $\left|r\right\rangle$ state 
via 2-photon resonance for $\Delta_c=50$ MHz. Hence, $\rho^{(0)}_{rr}$ in equations (14) and (15) are non-zero for zero velocity class
of atoms which interact with the weak probe beam and contribute to $\chi_{3L}(\omega_p+\delta)$.   
Since the strong probe beam resonantly interacts with a different velocity class of atoms, the 2-photon resonance for that velocity class 
is shifted due to wave vector mismatch and the corresponding EIT peak appears at $81.25$ MHz. To show that the above approximation is valid, 
we calculated $\chi_{3L}(\omega_p+\delta)$ using equations (14) and (15) which is shown in figure (\ref{fig1}b) and the approximation
holds very well.        
  
\begin{figure}[t]
\begin{center}
\epsfig{file=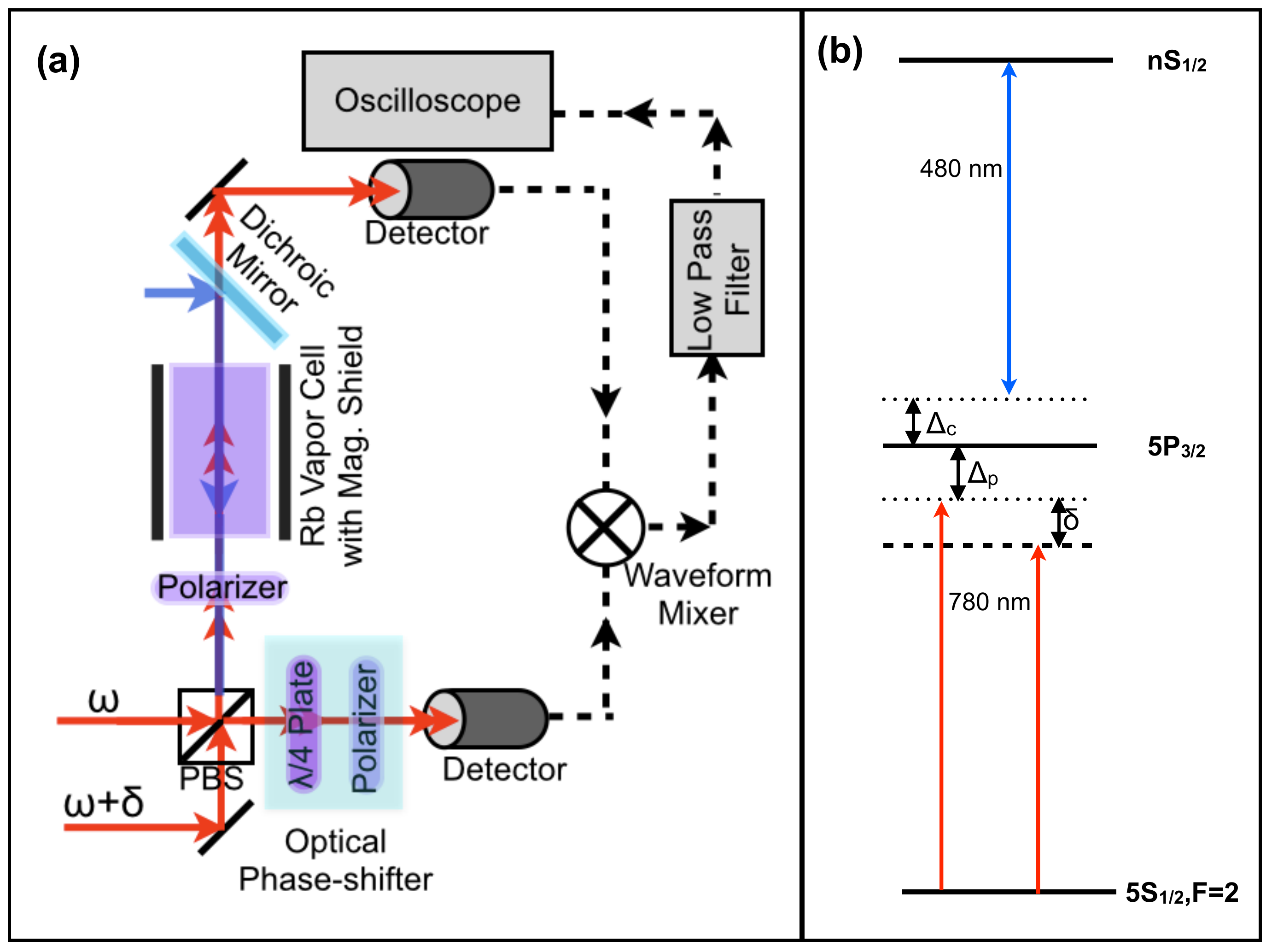,clip=,angle=0,width=8.0cm}
\caption[]{(a) Schematic of the experimental setup for heterdyne detection technique to measure the transmission and dispersion of Rydberg EIT 
medium. (b) Energy level diagram for Rydberg EIT in $^{87}$Rb. Two probe beams couple the transition $5$S$_{1/2}$, F$=2$ 
$\longrightarrow$ $5$P$_{3/2}$. The coupling laser couples the transition $5$P$_{3/2}$ $\longrightarrow$ 
$n$S$_{1/2}$. The coupling detuning is $\Delta_c$, probe detuning is $\Delta_p$ and frequency offset between the probe beams is $\delta$.}
\label{fig3}
\end{center}
\end{figure} 

Due to the wave vector mismatch in this case, the small peak is resolved from the EIT peak of the strong probe and a standard model for EIT with a single probe field and a coupling field can be used to compare with the experimental data. If the wave vectors are same, e.g. in the case of $\Lambda$ EIT in alkali atoms, the small peak can't be resolved from the EIT peak of the strong probe and hence, the model with two probes fields and a coupling 
field presented here should be used to compare with the experiment. Alternatively, the small peak can be reduced by changing the offset frequency. Using our model, the transmission peak height of the small peak is studied as a function of the offset frequency which is shown in figure~\ref{fig2}. It shows that the small peak height reduces significantly for higher offset frequency.

\section{Experimental Method}
The schematic of the experimental setup is shown in figure \ref{fig3}. An external cavity diode laser operating at 780 nm is used to derive two probe beams. A frequency offset of $50$ MHz was introduced between the probe beams by using acousto-optic modulators. Both the beams were superimposed using a polarizing cube beam splitter (PBS). The interference beat of the probes were detected using two fast photo-detectors by introducing polarizers at both the output ports of the PBS. The probe beams coming out of one of the output ports of the PBS propagate through a magnetically shielded rubidium vapor cell with optical path length of $5$ cm. The coupling beam was derived from a frequency doubled diode laser operating at $478 - 482$ nm and it counter-propagates the probe beams through the vapor cell.
The beat detected at the other output port of the PBS was used as reference. Since the frequency offset between the probe beams is larger than the Rydberg EIT resonance in thermal vapor~\cite{moha07}, they undergo different phase shift and absorption while scanning the coupling laser through the EIT resonance. This differential phase shift of the probe beams will change the phase of the signal beat which can be measured by comparing it with the phase of the reference beat. Since the beat signals are the outputs of the same interferometer, the noise due to vibration or acoustic disturbances are strongly suppressed.

\begin{figure}[t]
\begin{center}
\epsfig{file=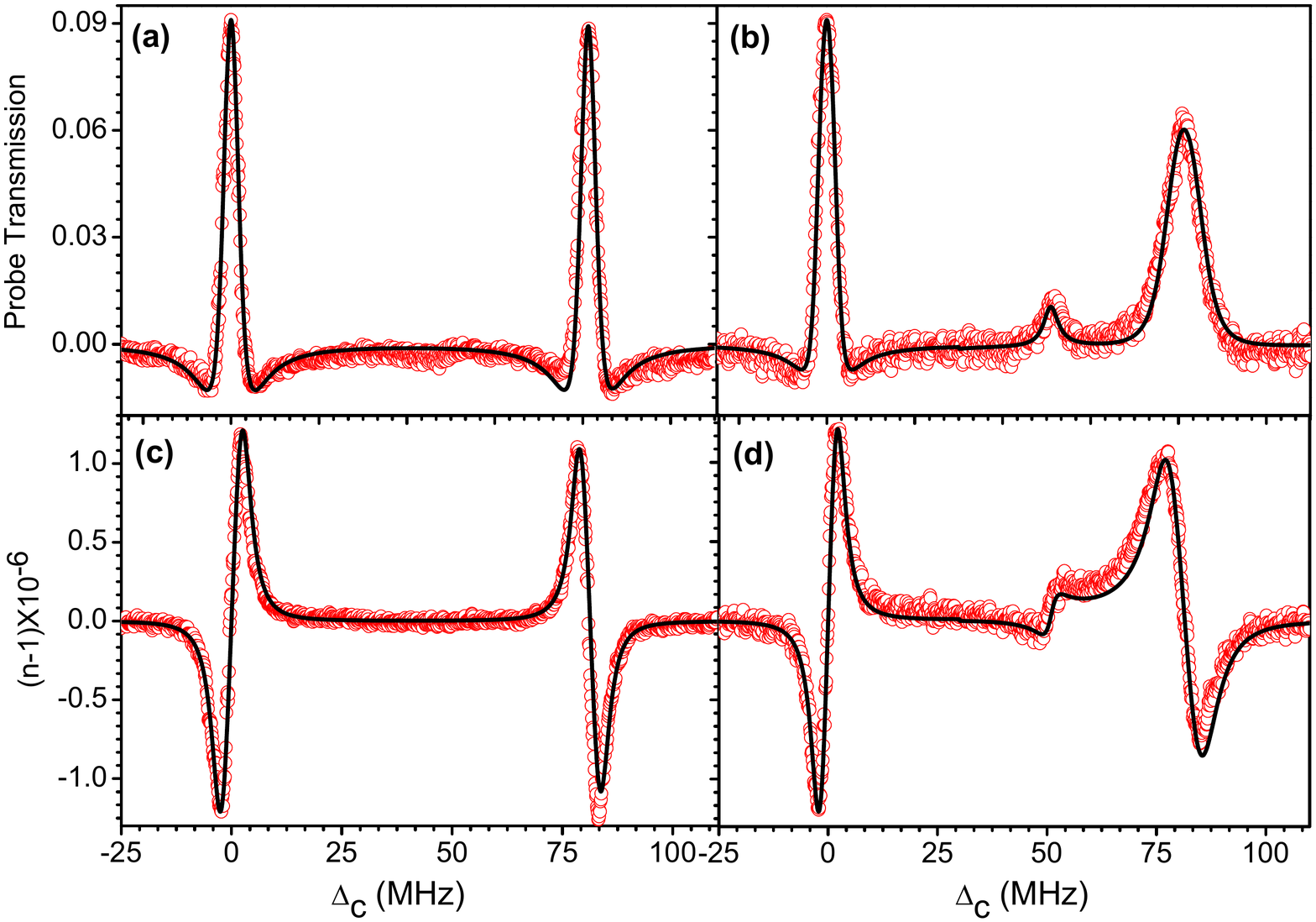,clip=,angle=0,width=8.0cm}
\caption[]{Absolute transmission and dispersion of the probe beams propagating through the Rydberg EIT medium after subtracting the offset due to their interaction with the $5$S$_{1/2}$, F$=2$ $\longrightarrow$ $5$P$_{3/2}$ transition in the absence of coupling beam. EIT (a) transmission and (c) dispersion signals of both the probes with Rabi frequency 600 kHz each. EIT (b) transmission and (d) dispersion signals of both the probes with Rabi frequencies, 600 kHz and 6.5 MHz respectively. The coupling Rabi frequency was 2.5 MHz in all the cases. The red open circles are experimental data points and the black solid lines are the curves generated by the model. The absolute transmission and dispersion were determined from the theoretical 
calculation for the given experimental parameters.}
\label{fig4}
\end{center}
\end{figure}

The light intensity falling at the signal detector is 
\begin{eqnarray}
I_s\propto && \left|E_1\right|^2e^{-klIm\left[\chi\left(\omega\right)\right]}+
\left|E_2^2\right|e^{-klIm\left[\chi\left(\omega+\delta\right)\right]} \nonumber \\
&&+2\left|E_1\right|\left|E_2\right|e^{-\frac{kl}{2}Im\left[\chi\left(\omega\right)
+\chi\left(\omega+\delta\right)\right]}\cos\left(\delta t+\phi_s+\phi_{off}\right) \nonumber 
\end{eqnarray}
where $E_1$ and $E_2$ are the electric field amplitudes of the strong and
weak probes respectively, $\phi_s=\frac{kl}{2}Re\left[\chi_{3L}\left(\omega\right)-\chi_{3L}\left(\omega+\delta\right)\right]$ and 
$\phi_{off}$ is the phase difference of the probe beams in the absence of the coupling field which remains constant if the probe frequencies 
are kept constant during the experiment. Using a high pass filter, the beat signal detected by the signal detector has the form
\begin{eqnarray}
D_s=A_se^{-\frac{kl}{2}Im\left[\chi_{3L}\left(\omega\right)
+\chi_{3L}\left(\omega+\delta\right)\right]}\cos(\delta t+\phi_s+\phi_{off}) \nonumber
\end{eqnarray}
where $A_s\propto 2\left|E_1\right|\left|E_2\right|$. Similarly, the beat signal at the reference detector has the form $D_r=A_r\cos(\delta t+\phi_r)$, where $A_r$ and $\phi_r$ are the amplitude 
and phase of the beat signal of the reference detector. $\phi_{r}$ can be controlled using an external phase shifter. These two beat signals 
are multiplied by an electronic waveform mixer and are passed through a low pass filter. The output of the low pass filter gives a DC signal 
of the form
\begin{eqnarray}
S_{L}=2A_{r}A_{s}e^{-\frac{kl}{2}Im\left[\chi_{3L}\left(\omega\right)+\chi_{3L}\left(\omega+\delta\right)\right]}\cos\left(\phi_{s}+\phi_{0}\right) \nonumber
\end{eqnarray}
where $\phi_0=\phi_r+\phi_{off}$. Assuming $\phi_s$ to be small and setting $\phi_{0}=0$, the signal becomes sensitive to the amplitudes of the 
probe beams and hence, gives the information about the transmission of the probe beams through the medium. After subtracting the offset (in absence of coupling laser) from the signal,
\begin{eqnarray}
S_L\approx 2A_sA_r\left[e^{-\frac{kl}{2}Im\left[\chi_{3L}\left(\omega\right)+\chi_{3L}\left(\omega+\delta\right)\right]}-1\right]
\end{eqnarray}
If $\phi_0$ is set to $\frac{\pi}{2}$, then $S_L$ becomes strongly sensitive to $\phi_s$ and and hence, the refractive index of the probe beams 
due to Rydberg EIT can be measured. In this case, 
\begin{eqnarray}
S_L\approx 2A_sA_re^{-\frac{kl}{2}Im\left[\chi_{3L}\left(\omega\right)+\chi_{3L}\left(\omega+\delta\right)\right]}\phi_s
\end{eqnarray} 
It is worthwhile to mention that, the observed dispersive signal depends linearly on $\phi_s$ and hence, proportional to 
$[Re(\chi_{3L} (\omega))-Re(\chi_{3L} (\omega+\delta))]$. 

To work in the phase as well as amplitude sensitive regimes, $\phi_{0}$ can be controlled by varying the phase of the reference beat signal using electronic phase shifter. However, in our experiment, the phase is controlled optically and we call it as optical phase shifter (OPS). To realize the OPS, one $\frac{\lambda}{4}$-plate is introduced before the polarizer at the output of the PBS. The probe beam transmitted (reflected) by the PBS after passing through the $\frac{\lambda}{4}$-plate become $\sigma^{+}$ ($\sigma^{-}$) and can be expressed as $\sigma^{\pm}=\frac{1}{\sqrt{2}} (|H\rangle + e^{\pm i\frac{\pi}{2}}|V\rangle)$. If the polarizer after $\frac{\lambda}{4}$-plate selects the $|H\rangle$-polarized component, then the phase difference between both the probe beams falling on the detector is zero. Now, if the angle of the polarizer axis is rotated by $90^{\circ}$, then $|V\rangle$ is selected and the phase difference between the beams becomes $\pi$. Hence, by rotating the polarizer axis, the phase of the reference beat signal can be varied between $0$ to $\pi$ without compromising the amplitude.

\begin{figure}[t]
\begin{center}
\epsfig{file=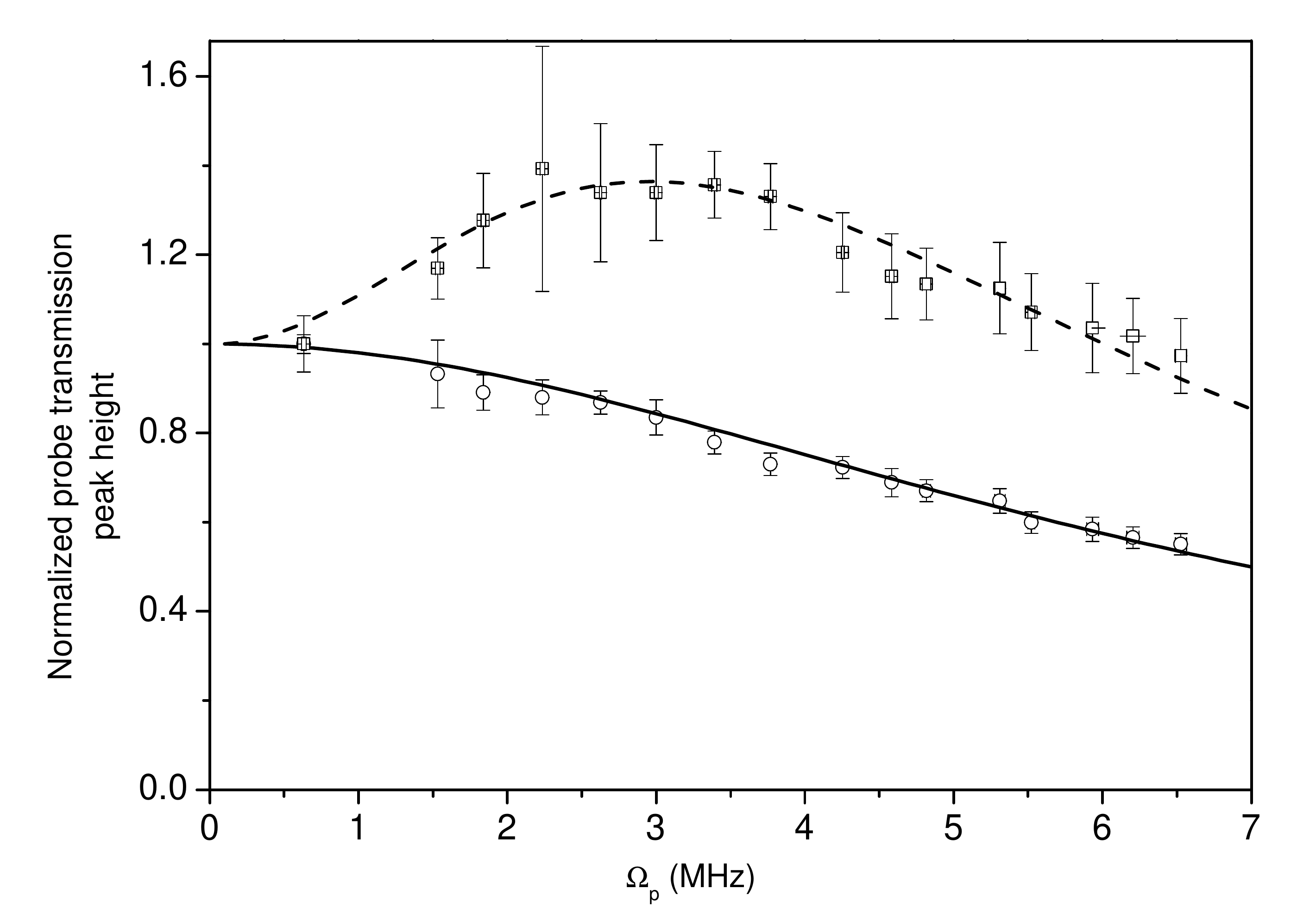,clip=,angle=0,width=8.0cm}
\caption[]{Normalized EIT transmission peak height as a function of probe Rabi frequency by keeping coupling Rabi frequency fixed at $2.5$ MHz $({\circ})$ and $800$ kHz $({\square})$. The solid and dashed lines are the curves generated using the model for same experimental parameters.}
\end{center}
\label{fig5}
\end{figure} 

The typical transmission and dispersion signals of the probes propagating through the Rydberg EIT medium are shown in figure \ref{fig4}. The probe laser frequency was stabilized on the atomic transition ($5$S$_{1/2}$, F$=2$ $\longrightarrow$ $5$P$_{3/2}$)  of $^{87}$Rb and the coupling laser frequency was scanned through the Rydberg EIT resonance. The frequency offset between the probe beams is $50$ MHz which is much greater than the EIT resonance width (about 3 MHz in thermal vapor~\cite{moha07}). Hence, two distinct EIT transmission peaks are observed as shown in figure \ref{fig4}(a) when the phase difference between the beat signals was set to zero. When the phase difference was set to $\pi/2$, two respective dispersion signals were observed as shown in figure \ref{fig4}(c). Due to the wave vector mismatch between the probe and the coupling beams, the frequency difference between the transmission or dispersion peaks respective to both the probe beams are scaled by $\frac{k_c}{k_p}$ as discussed in section II. We observe the frequency difference between these peaks to be $81.25$ MHz which is consistent with the above scaling. With the increased Rabi frequency of the strong probe, the small peak at the beat frequency appears as discussed in section II and shown in figure~\ref{fig4}(b) and~\ref{fig4}(d). The $1/e^{2}$-radius of probe (coupling) is measured to be $0.7$ mm ($1.2$ mm). The power of the weak probe, used in the experiment was $0.125$ $\mu$W. The strong probe power was varied in the range of $0.125$ $\mu$W to $15$ $\mu$W. Probe Rabi frequency is estimated as, $\Omega_{p}=\Gamma_{eg} 
\sqrt{\frac{I}{2I_{sat}}}$. For $^{87}$Rb, the saturation power is $I_{sat}=1.64$ mW/cm$^{2}$, and lifetime of $^{5}$P$_{3/2}$ state is 
$\Gamma_{eg}= 6$ MHz. The coupling Rabi frequency is determined by fitting the EIT  transmission peak for a weak probe beam.

\section{Measurement of optical nonlinearity}
In order to study the optical non-linearity, the weak probe Rabi frequency was set to 600 kHz and the strong probe Rabi frequency is varied 
from 600 kHz to 6.5 MHz. The EIT peak height of the weak probe is used as reference to normalize the EIT peak height of the strong probe beam. 
The beat frequency is chosen sufficiently large such that the peak due to the dressed atoms is well resolved from the main EIT peak of the strong 
probe beam. The normalized transmission peak height of the probe beam as a function of its Rabi frequency is shown in figure~\ref{fig5}. The curves generated using the above model fit well with the experimental data as shown in figure~\ref{fig5}. In this particular case, since the main EIT peak is 
well resolved from the peak due to the dressed atom, the standard EIT model using a single probe beam also fits well with the peak height data and shows very little deviation from our model. 

\begin{figure}[t]
\begin{center}
\epsfig{file=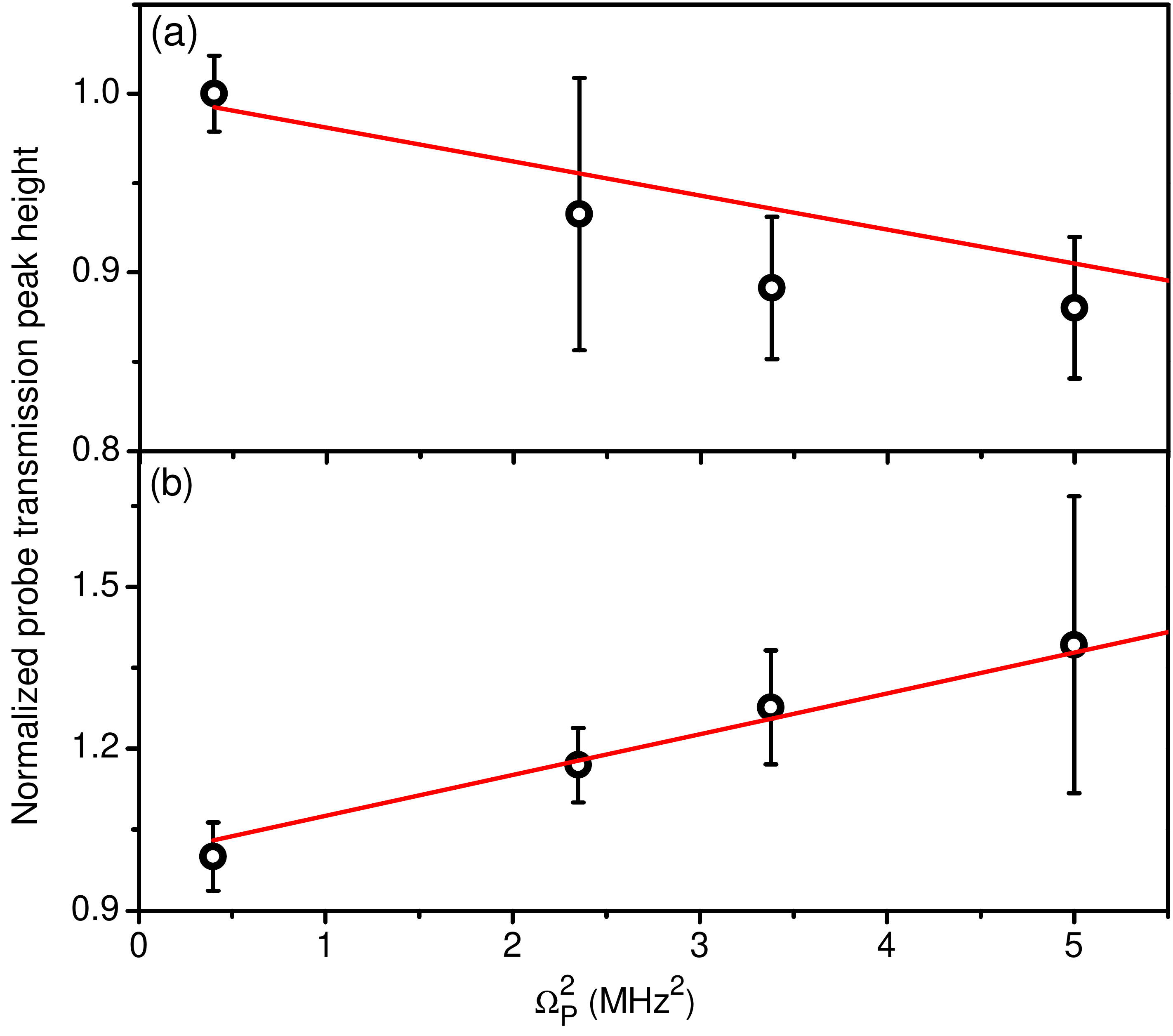,clip=,angle=0,width=8.0cm}
\caption[]{Normalized EIT peak height as a function of $\Omega_p^2$. The circles are the experimental data with 
(a) $\Omega_c=2.5$ MHz and (b) $\Omega_c=0.8$ MHz. The solid lines are the linear fitting with the function $1+a\Omega_p^2$ with $a$ as 
the fitting parameter.}
\label{fig6}
\end{center}
\end{figure} 

To determine the contributions of the higher order susceptibilities to EIT peak, we do the following analysis.\\ 
The EIT peak height of weak probe is given by,
\begin{eqnarray}
P_w&&=S_L(\Delta_c=0) \nonumber \\
&&=2A_sA_r\left[e^{-\frac{kl}{2}Im\left[\chi_{3L}\left(\omega+\delta\right)\right]}-1\right] \nonumber
\end{eqnarray} 
Similarly the EIT peak height of the strong probe beam is given by
\begin{eqnarray}
P_s&&=S_L(\Delta_c=\frac{k_c}{k_p}\delta) \nonumber \\
&&=2A_sA_r\left[e^{-\frac{kl}{2}Im\left[\chi_{3L}\left(\omega\right)\right]}-1\right] \nonumber
\end{eqnarray} 
Assuming that $\chi_{3L}(\omega)$ and $\chi_{3L}(\omega+\delta)$ are small, the ratio is
$\frac{P_s}{P_w}\approx\frac{Im\left[\chi_{3L}(\omega)\right]}{Im\left[\chi_{3L}(\omega+\delta)\right]}$. 

The Taylor expansion of the susceptibility is given by 
\begin{eqnarray}
\chi_{3L}&&=\left[\chi_{3L}\right]_{\Omega_{p}=0} 
+\frac{1}{2!}\left[\frac{\partial^{2}\chi_{3L}}{\partial \Omega^{2}_{p}}\right]_{\Omega_{p}=0}\Omega^{2}_{p} \nonumber \\
&&+\frac{1}{4!}\left[\frac{\partial^{4} \chi_{3L}}{\partial\Omega^{4}_{p}}\right]_{\Omega_{p}=0} \Omega^{4}_{p}+...\nonumber
\end{eqnarray}
Since $\chi_{3L}$ is an even function of $\Omega_P$, all the odd order terms in the expansion are zero. $\chi_{3L}$ can be expressed as 
\begin{eqnarray}
\chi_{3L}&=&\chi^{(1)}_{3L}+\chi^{(3)}_{3L}E_p^2+\chi^{(5)}_{3L}E_p^{4}+...\nonumber\\
&=&\chi^{(1)}_{3L}+\chi^{(3)}_{3L}[\frac{\hbar}{2\mu}]^{2} \Omega_p^2+\chi^{(5)}_{3L}[\frac{\hbar}{2\mu}]^{4} \Omega_p^{4}+...\nonumber
\end{eqnarray}

Where $E_{p}$ is the probe electric field. Comparing both the equations, we get $\chi^{(1)}_{3L}=\chi_{3L}(\Omega_{p}=0)$, $\chi^{(3)}_{3L}=\frac{1}{2!} [\frac{2\mu}{\hbar}]^{2} {\left[\frac{\partial^{2}\chi_{3L}}{\partial \Omega^{2}_{p}}\right]}_{\Omega_{p}=0}$, $\chi^{(5)}_{3L}=\frac{1}{4!} [\frac{2\mu}{\hbar}]^{4}{\left[\frac{\partial^{4}\chi_{3L}}{\partial \Omega^{4}_{p}}\right]}_{\Omega_{p}=0}$. Since the Doppler broadening is much larger than the offset frequency $\delta$, it is assumed that $\chi^{(1)}_{3L}(\omega)\approx\chi^{(1)}_{3L}(\omega+\delta)$. Also for the weak probe beam, the higher order terms are assumed to be negligible. So the normalized EIT peak height of the strong probe beam can be written as
\begin{eqnarray}
\frac{P_{s}}{P_{w}}=1+\frac{Im(\chi^{(3)}_{3L})}{Im(\chi^{(1)}_{3L})} \Omega^{2}_{p} + \frac{Im(\chi^{(5)}_{3L})}{Im(\chi^{(1)}_{3L})} 
\Omega^{4}_{p}+...\nonumber
\end{eqnarray}

In principle, the above polynomial function can be used to fit the transmission peak height data as shown in figure~\ref{fig5} to determine the 
higher order non-linearities. Though the exact solution of the EIT fits the data very well, keeping a few terms in the above polynomial function doesn't 
fit the data equally well mainly due to large contributions of the higher order terms at higher probe Rabi frequencies. Therefore, we selected
the first four data points of figure~\ref{fig5} to fit with a function $1+a\Omega_p^2$, where $a=\frac{Im(\chi^{(3)}_{3L})}{Im(\chi^{(1)}_{3L})}$ 
and gives information about the self phase modulation ($\chi^{(3)}_{3L}$) of the probe light. From the fitting, we find the value of "$a$" to be $-0.02\pm0.004$ MHz$^{-2}$ and  $0.076\pm0.006$ MHz$^{-2}$ with coupling Rabi frequencies $2.5$ MHz and $0.8$ MHz respectively. To compare with the theory, $\chi^{(3)}_{3L}=\frac{1}{2!}[\frac{2\mu}{\hbar}]^{2} {\left[\frac{\partial^{2}\chi_{3L}}{\partial \Omega^{2}_{p}}\right]}_{\Omega_{p}=0}$ was calculated using the same experimental parameters and the value of "$a$" was found to be $-0.014$ MHz$^{-2}$ and $0.064$ MHz$^{-2}$ with coupling Rabi frequencies equal to $2.5$ MHz and $0.8$ MHz respectively. $\chi^{(3)}_{3L}$ determined using above analysis reasonably match with the theoretical calculation. The discrepancy is mainly due to the non-zero contribution of higher order terms. More number of data points below $1$ MHz may give a better measurement for $\chi^{(3)}_{3L}$. Higher order non-linearity can not be determined accurately as the series diverges very fast by increasing the probe Rabi frequency for this system.   

\section{Conclusion}
We have demonstrated a technique based on optical heterodyne and presented a suitable model to measure the optical non-linearity (self phase modulation) of a probe beam propagating through a dispersive medium accurately. The technique can also be used to measure the cross phase modulation of the light field propagating through a highly dispersive medium. Recently, the technique has been used to demonstrate the blockade in two-photon excitations to the Rydberg state in thermal vapor~\cite{bhow16}. We would like to extend this technique to measure the optical non-linearity of Rydberg EIT in blockade interaction regime in thermal vapor as well as in ultra-cold atoms.

\section{Acknowledgment}
We acknowledge Dushmanta Kara for assisting in performing the experiment. This experiment was financially supported by the Department of 
Atomic Energy, Govt. of India.

\end{document}